\documentclass[12pt]{aastex}

\newcommand\iso[2]{${\rm ^{#2}}$#1}

\def\BmV0{\mbox{(B-V)$^{\rm o}$}}
\def\VmK0{\mbox{(V-K)$^{\rm o}$}}
\def\MV0{\mbox{M$_{\rm V}^{\rm o}$}}
\def\MV{\mbox{M$_{\rm V}$}}

\def\deg{{$^{\circ}$}}
\def\bd17{\mbox{BD +17\deg 3248}}

\begin{document}

\received{}
\accepted{}

\title{Europium Isotopic Abundances in Very Metal-poor Stars\altaffilmark{1}}

\author{
Christopher Sneden\altaffilmark{2}, 
John J. Cowan\altaffilmark{3},
James E. Lawler\altaffilmark{4},
Scott Burles\altaffilmark{5},
Timothy C. Beers\altaffilmark{6}, and
George M. Fuller\altaffilmark{7} \\
}

\altaffiltext{1}{Based on observations obtained with: (a) the Keck~I 
Telescope of the W. M. Keck Observatory, which is operated by the 
California Association for Research In Astronomy (CARA, Inc) on 
behalf of the University of California and the California Institute 
of Technology: and (b) the H. J. Smith Telescope of McDonald Observatory,
which is operated by the University of Texas at Austin.}
  
\altaffiltext{2}{Department of Astronomy and McDonald Observatory,
University of Texas, Austin, TX 78712; chris@verdi.as.utexas.edu}

\altaffiltext{3}{Department of Physics and Astronomy,
University of Oklahoma, Norman, OK 73019; cowan@phyast.nhn.ou.edu}

\altaffiltext{4}{Department of Physics, University of Wisconsin, 
Madison, WI 53706; jelawler@facstaff.wisc.edu}

\altaffiltext{5}{Department of Physics, Massachusetts Institute of Technology,
77 Massachusetts Avenue, Room 6-113, Cambridge, MA 02139-4307; burles@mit.edu}
 
\altaffiltext{6}{Department of Physics and Astronomy, Michigan State 
University, East Lansing, MI 48824; beers@pa.msu.edu}
 
\altaffiltext{7}{Department of Physics, University of California at 
San Diego, La Jolla, CA 92093-0319; gfuller@ucsd.edu}

\slugcomment{Submitted to {\it The Astrophysical Journal Letters}}

\begin{abstract}

Europium isotopic abundance fractions are reported for the very metal-poor,
neutron-capture-rich giant stars CS~22892-052, HD~115444, and \bd17.
The abundance fractions, derived from analysis of several strong
\ion{Eu}{2} lines appearing in high-resolution spectra of these stars,
are in excellent agreement with each other and with their values in
the Solar System: 
fr(\iso{Eu}{151})~$\simeq$ fr(\iso{Eu}{153})~$\simeq$ 0.5.
Detailed abundance studies of very metal-poor stars have previously 
shown that the total elemental abundances of stable atoms with atomic 
numbers z~$\geq$~56 typically match very closely those of a scaled 
solar-system $r$-process abundance distribution. 
The present results for the first time extend this agreement to 
the isotopic level.

\end{abstract}


\section{Introduction}

Neutron bombardment reactions create all abundant isotopes of more than 
60 elements of the periodic table.
Neutron-capture ($n$-capture) nucleosynthesis can occur slowly in the
$s$-process, in which $\beta$ decays have time to happen between successive
neutron captures, or rapidly in the $r$-process, in which enormous,
but short-lived, neutron fluxes temporarily overwhelm $\beta$-decay rates.
After completion of an $n$-capture event, the final isotopic abundance
mix will be very different for the $r$- and $s$-processes.
Detailed exploration of these $n$-capture events can best be done 
with full knowledge of the isotopic abundances in stars over a large
metallicity range. 
At present, this can be done for one star, the Sun, through the isotopic 
analysis of carbonaceous chondrite meteorites.  
Such analysis reveals a complex mix of $r$- and $s$-process 
contributions to Solar System $n$-capture elements.

Meteoritic data can be obtained only for the Solar System, so $n$-capture 
contents of other stars must be gleaned from their spectra.
However, the isotopic wavelength shifts for transitions of almost all 
$n$-capture elements are small compared to the thermal+turbulent line 
widths in stellar spectra. 
Thus isotopic abundance information is usually inaccessible, and 
confrontation of $n$-capture predictions can only be 
done with total elemental abundances.
In the most metal-poor stars, these abundances are very different 
than their total ($s$- plus $r$-process) Solar System values. 

Neglecting the cases of anomalously carbon-rich stars, abundances of
$n$-capture elements with Z~$\geq$ 56 in extremely metal-poor stars 
are consistent only with the $r$-process component abundances in 
Solar-System material (e.g. McWilliam 1997\nocite{McW97}; Burris et~al.  
2000\nocite{BPASCR00}; Sneden et~al. 2000\nocite{SCIFBBL00};
Westin et~al 2000\nocite{WSGC00}; Cayrel et~al. 2001\nocite{Cay01}).
This agreement suggests that even though the $r$-process is a
very violent dynamical event that cannot be reproduced in the laboratory,
and is very difficult to model theoretically, nature may be constrained
to produce $r$-process abundances in basically one fashion.
But the case could be strengthened if the elemental abundances
could be deconvolved into isotopic fractions.
In this {\em Letter} we report a step in this direction, by demonstrating
that abundance fractions of europium's two stable isotopes are essentially
equal in three $n$-capture-rich halo stars, just as they are in the 
Solar System.

\section{Observational Data and Isotopic Abundance Analysis}

We consider three very metal-poor, but relatively $n$-capture-rich, 
giant stars for which high-resolution (R~$\equiv$~$\lambda/\Delta\lambda$~=
45,000--60,000), high signal-to-noise (S/N~$>$ 100 in the wavelength 
regions of interest) spectra have been employed in separate chemical
composition analyses.
The stars are: 
(a) CS~22892-052 (Keck~I data, R~$\simeq$ 45,000; [Fe/H]~$\simeq$ --3.1 and 
[Eu/Fe]~$\simeq$ +1.6, Sneden et~al. 2000\nocite{SCIFBBL00});
(b) HD~115444 (McDonald data, R~$\simeq$ 60,000; [Fe/H]~$\simeq$ --2.9 and 
[Eu/Fe]~$\simeq$ +0.8, Westin et~al. 2000\nocite{WSGC00}); and
(c) \bd17\ (Keck and McDonald data, [Fe/H]~$\simeq$ --2.1 and 
[Eu/Fe]~$\simeq$ +0.9, Cowan et~al. 2001\nocite{Cowetal01}).
We adopted the model atmosphere parameters and elemental abundances
determined in these studies, which the reader should consult for
detailed discussion of spectrum acquisition, data reduction, and
overall abundance analysis.

In cool giant stars, europium exists almost exclusively in the \ion{Eu}{2}
species.
About a half-dozen spectroscopically accessible strong transitions 
from low-lying excitation states of this ion can be studied. 
The \ion{Eu}{2} energy levels have substantial hyperfine substructure,
and the isotopic energy level shifts between stable isotopes \iso{Eu}{151} and 
\iso{Eu}{153} are also large, creating complex and broad absorption features.  
The ground-state configuration of \ion{Eu}{2} is 
4$f^7$($^8$S)6$s$, from which arise all of the strong blue-violet 
transitions that can be detected in stellar spectra.
The unpaired 6$s$ electron of this configuration produces large magnetic 
hyperfine splitting, making the blue-UV lines intrinsically broad.
The 6$s$ electron also produces the large isotope shift.  
The isotope shift is primarily a field shift due to the finite, 
and different, size of the two Eu nuclei.

Accurate new laboratory $gf$, hyperfine structure, and isotopic shift data
for \ion{Eu}{2} have been determined by Lawler, Bonvallet, \& Sneden 
(2001)\nocite{LBS01}.
In their application of these data to the solar spectrum, Lawler et~al.
used synthetic spectrum computations to demonstrate (following the early 
work of Hauge 1972\nocite{Hau72}) that the Eu abundance fractions are 
essentially equal: fr(\iso{Eu}{151})~$\simeq$ fr(\iso{Eu}{153})~$\simeq$ 0.5, 
in good (and expected) agreement with the meteoritic values recommended
by Anders \& Grevesse (1989)\nocite{AG89}: fr(\iso{Eu}{151})~= 0.478, 
fr(\iso{Eu}{153})~= 0.522.

We adopted without change the line lists of Lawler et~al. (2001), which 
include not only the \ion{Eu}{2} hyperfine/isotopic components, but other 
atomic and molecular lines in small spectral regions surrounding the 
Eu lines of interest.
Initial synthetic spectrum computations showed that the \ion{Eu}{2} lines
at 3819.67, 3907.11, 4129.72, and 4205.04~\AA\ have the best combination
of large isotopic splitting and relative freedom from contamination
from other spectral absorption features.
Then for each of these lines in each program star, we iteratively determined 
the best Eu isotopic fractions in the following manner. 
First, the line profiles over the whole synthesized wavelength region 
were used to determine the optimum spectral smoothing (a combination 
of spectrograph resolution and so-called stellar macroturbulent broadening).
Next, as a trial we assumed that the meteoritic Eu isotopic fractions 
were correct, and determined the best Eu elemental abundance for the 
\ion{Eu}{2} line (which always turned out to be within $\pm$0.03~dex of the 
mean Eu abundance for the star).
Finally, the elemental abundance was fixed and the observed line profile
was compared to synthetic profiles with different assumed isotopic fractions.
This procedure was repeated until the spectral line shape was reproduced
satisfactorily.

As one check of the isotopic fractions derived in this manner, we also
determined the best overall synthetic/observed spectrum matches with
simultaneous variations in elemental and isotopic abundances in 
producing the minimum residuals in the spectrum fits.  
However, the derived isotopic fractions proved to be essentially the
same as those determined in the simpler procedure described in the 
previous paragraph.

\begin{figure} 
\epsscale{0.80}
\plotone{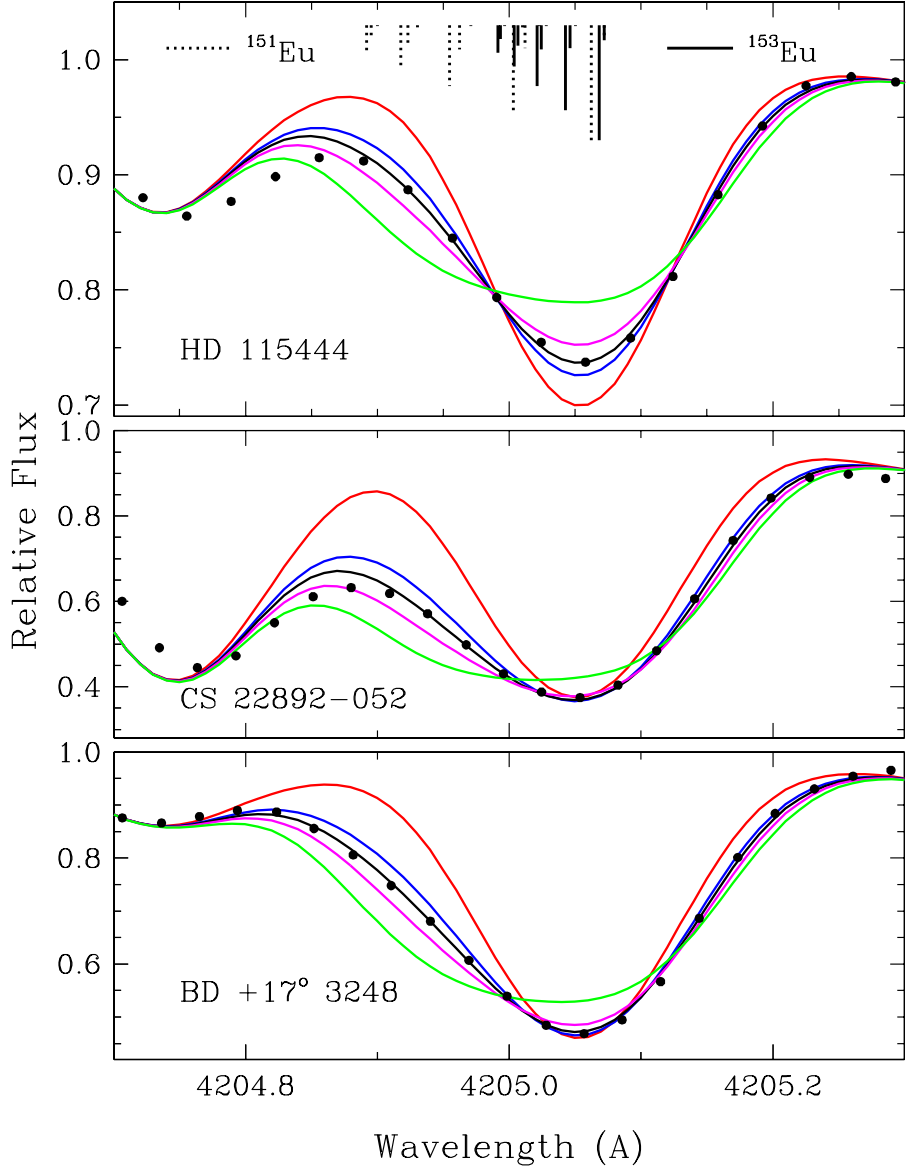}
\caption{Observed and synthetic spectra of the \ion{Eu}{2} 4205.04~\AA\
line in the three program stars.
The observed spectra (filled circles) are compared to synthetic spectra 
with fr(\iso{Eu}{151}) = 0.000 (red line); 0.350 (blue line); 
0.478 (black line); 0.650 (magenta line); and 1.000 (green line).
As Eu has only two stable isotopes, fr(\iso{Eu}{153})~= 
1.0~--~fr(\iso{Eu}{151}).
In the top panel vertical lines are added to indicate the wavelengths
and relative strengths of the hyperfine components of the isotopes
\iso{Eu}{151} (dotted lines) and \iso{Eu}{153} (solid lines).  
The absolute vertical line lengths are normalized by an arbitrary 
constant for display purposes.
\label{spec4205}}
\end{figure}

\begin{figure} 
\epsscale{0.80}
\plotone{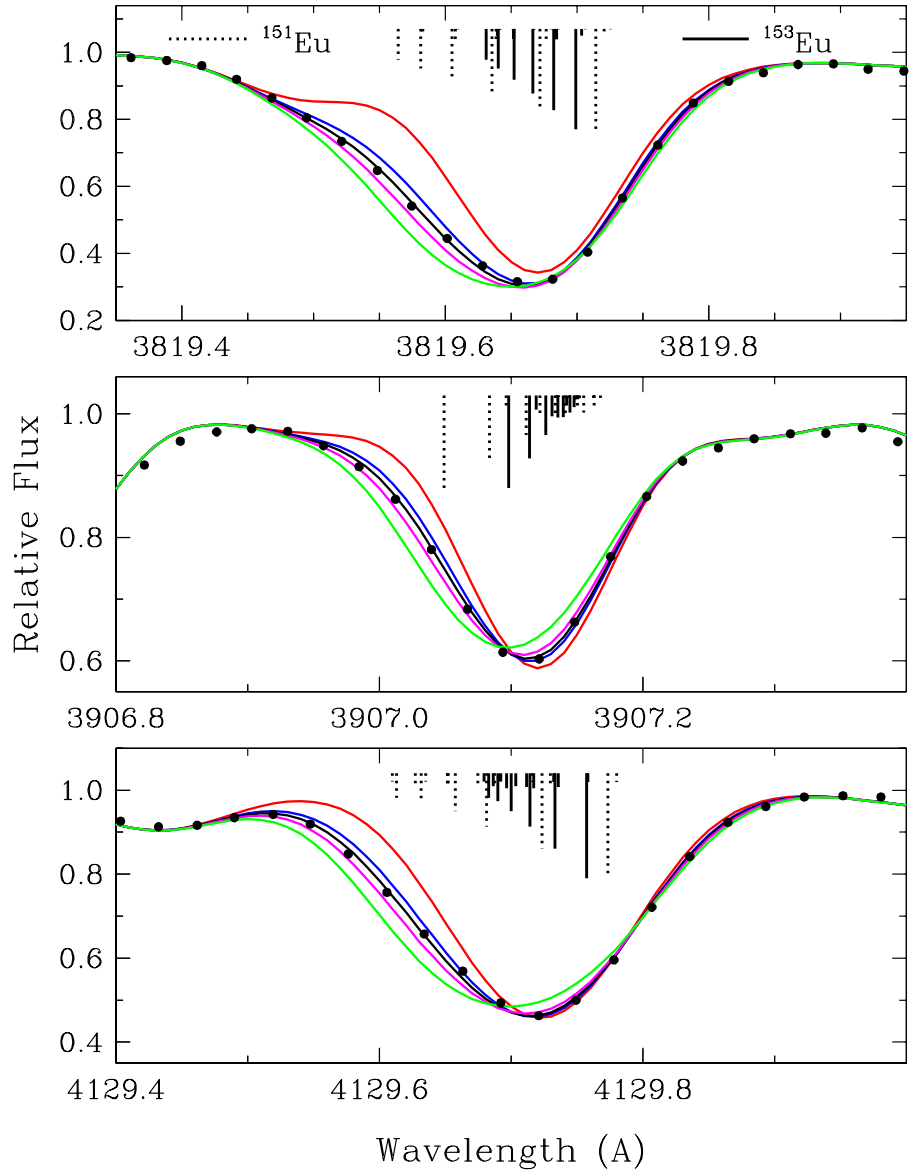}
\caption{Observed and synthetic spectra of the \ion{Eu}{2} 3907.11,
4129.72, and 4205.04~\AA\ lines in \bd17.
Lines and symbols have the same meaning as those of Figure~1.
\label{specbd17}}
\end{figure}

In Figure~\ref{spec4205} we present synthetic and observed spectra
of the \ion{Eu}{2} 4205.04~\AA\ line in each of the three program stars. 
In the top panel of this figure we have also marked vertical lines to 
indicate the wavelengths and relative strengths of the \iso{Eu}{151} and 
\iso{Eu}{151} subcomponents for the \ion{Eu}{2} features.
In Figure~\ref{specbd17} the synthetic and observed spectra of
\bd17\ are shown for the other three \ion{Eu}{2} lines that we employed 
in isotopic abundance derivations.
In each panel of this figure the transition subcomponents are marked with 
vertical lines in the manner of Figure~\ref{spec4205}.
Inspection of these component distributions reveals that the hyperfine
pattern for \iso{Eu}{151} is much broader than that of \iso{Eu}{153},
in addition to being displaced blueward on average.
These circumstances considerably aid the derivation of isotopic fractions.
It is apparent from the \ion{Eu}{2} line profiles shown in these figures
that the assumption either of \iso{Eu}{151} or \iso{Eu}{153} dominance in
the isotopic mix will not work, and in fact the observed profiles seem
best matched by fr(\iso{Eu}{151})~$\simeq$ fr(\iso{Eu}{153})~$\simeq$ 0.5,
just as in the solar photospheric spectrum.

\begin{figure} 
\epsscale{0.80}
\plotone{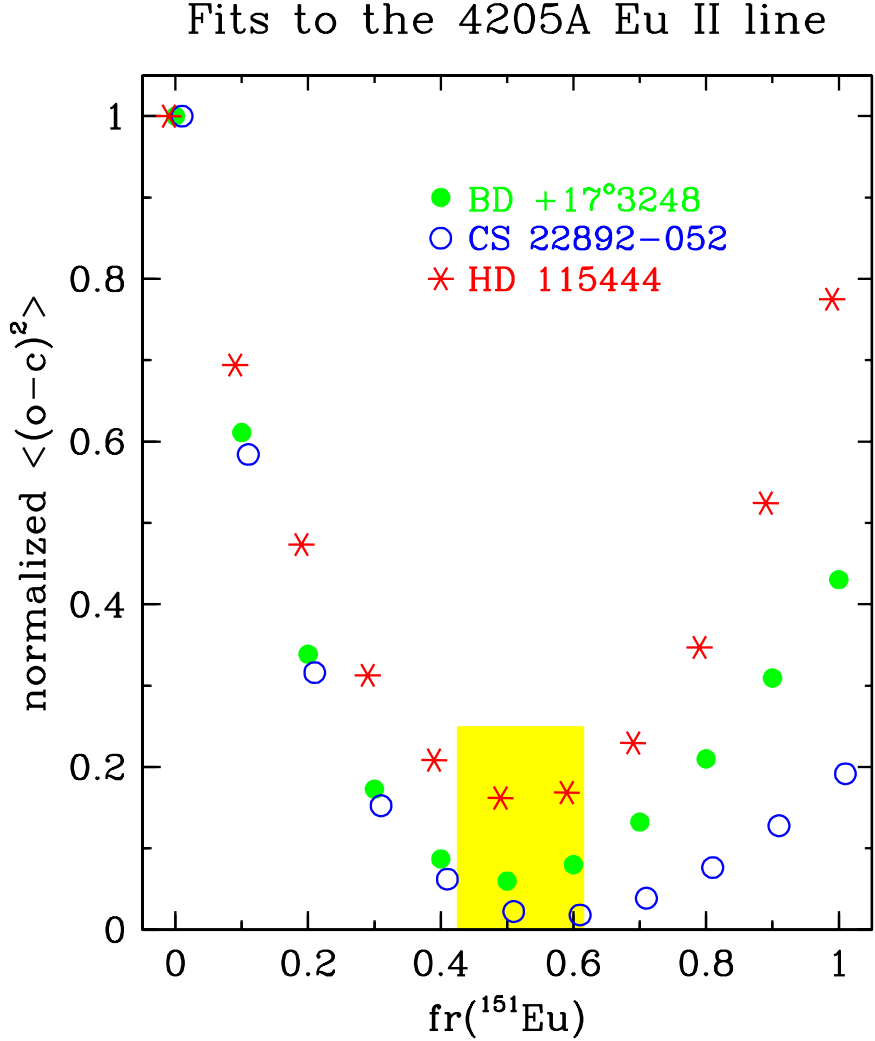}
\caption{Mean values of the squares of the deviations $(o-c)$
between observed and synthesized spectra of the 4205~\AA\ \ion{Eu}{2}
line in the three stars, plotted as functions of assumed fractional
\iso{Eu}{151} contributions.
Small horizontal shifts between the three sets of numbers have
been introduced for clarity of display.
Likewise, the data for the three stars have been normalized for display by
multiplying by a constant to force agreement at $<$(o-c)$^2>$ of the 
points at fr(\iso{Eu}{151})~= 0.0
\label{stats}}
\end{figure}

To obtain a more quantitative assessment of the isotopic fractions,
we computed goodness-of-fit parameters $<(o-c)^2>$ for the observed
4205~\AA\ line profiles under different isotopic fraction assumptions.
Here an $o-c$ deviation has been computed at each of the observed spectrum 
points (refer back to the filled circles displayed in 
Figure~\ref{spec4205}) as the observed spectrum depth 
$minus$ the computed depth.
These mean deviations between observed and synthetic spectra are plotted
in Figure~\ref{stats}, and they confirm the visual impressions 
from Figures~\ref{spec4205} and~\ref{specbd17}.
The analyses of each of the four \ion{Eu}{2} lines in the three program 
stars yield essentially identical results, hence we suggest that in 
very-metal-poor $r$-process-enhanced stars of the Galactic halo, 
fr(\iso{Eu}{151})~= fr(\iso{Eu}{153})~= 0.5~$\pm$~0.1. 
In short, we discern no deviation from the Eu isotopic mix in the 
Solar System.

\section{Discussion and Conclusions}

Both europium isotopes, \iso{Eu}{151}\ and \iso{Eu}{153}.
are produced primarily by the {\it r}-process in Solar System material
(see K{\"a}ppeler et al. 1989), and thus this element has only a very small
{\it s}-process contribution (Burris et~al. 2000\nocite{BPASCR00}).
These conclusions are based upon employing the {\it classical} or
{\it standard} model for predicting {\it s}-process production and the 
associated {\it r}-process Solar System residuals.
More complex models have been developed to predict {\it s}-process
abundances in low- to intermediate-mass asymptotic giant branch (AGB) stars, 
the likely dominant site for this type of synthesis.
However, even in these stellar models the calculated {\it s}-process abundances
for \iso{Eu}{151}\ and \iso{Eu}{153} are small, and the abundance ratio remains
solar (see Arlandini et~al. 1999\nocite{AKWGLBS99}).

We can also rule out large amounts of $s$-process contribution to the 
europium abundances of our program stars, either by past mass transfer 
from nearby AGB stars or from internal nucleosynthesis.
Such events would yield large $s$-process contributions to a number 
of elements, not just to europium. 
Consider, for example, the neighboring elements barium and lanthanum, whose
synthesis in Solar System matter has been dominated by the $s$-process.
The mean observed abundance ratios of these elements with respect to 
europium for the program stars, from results of Sneden et~al. (2000),
Westin et~al. (2000), and Cowan et~al. (2001), are 
$<$log~$\epsilon$(Ba/Eu)$>$~$\simeq$ +1.0 and 
$<$log~$\epsilon$(La/Eu)$>$~$\simeq$ +0.2.
But in the Solar System these ($r$- plus $s$-process) abundance ratios 
are +1.8 and +0.7, respectively (Burris et~al. 2000\nocite{BPASCR00}).
Moreover, $s$-process computations by Malaney (1987\nocite{Mal87}) covering
a large range of neutron exposure parameters yield $minumum$ values for
the ratios of +2.5 and +1.3, respectively.
It cannot be simultaneously argued that the europium abundances in
our stars have significant $s$-process components while their Ba/Eu
and La/Eu elemental abundance ratios are more than an order of magnitude
smaller than in the Solar System or in $s$-process predictions.\footnote{
One abundance anomaly among the three target stars should be noted.
The carbon abundance of CS~22892-052 is very large ([C/Fe]~$\simeq$ +1.0:
McWilliam et~al. 1995\nocite{MPSS95}; Norris, Ryan, \& Beers 
1997\nocite{NRB97}) compared to that of the other two stars 
([C/Fe]~$<$ 0: Kraft et~al. 1982\nocite{Ketal82}; Westin et~al.
2000; Cowan et~al. 2001).
Carbon overabundances in some very metal-poor stars usually are accompanied 
by significant $s$-process abundance enhancements (e.g, Norris
et~al. 1997\nocite{NRB97}; Hill et~al. 2000\nocite{Hetal97}).
However, some carbon-rich low metallicity stars are known to be 
$s$-process-poor (Preston \& Sneden 2001\nocite{PS01}; 
Aoki et~al. 2001\nocite{ANRBA01}).
Probably the large carbon abundance and $n$-capture enhancements
of CS~22892-052 were generated in different nucleosynthesis events.}
 
One further consideration would be whether the {\it r}-process itself 
might not always produce Solar System isotopic abundance ratios for europium.
Those abundances are determined, for example, by the neutron number density
and the corresponding flow ``path'' for this synthesis. 
However. such a variation would also affect other isotopes, producing 
non-solar elemental abundance ratios -- something that is not observed
in very metal-poor halo stars.
Recent {\it r}-process models (e.g., Pfeiffer, Kratz, \& Thielemann
1997\nocite{PKT97}; Cowan et~al. 1999\nocite{CPKTSBTB99}) 
that reproduce the observed solar {\it r}-process elemental abundances in 
these stars also replicate the Solar System isotopic ratios for elements
such as europium.
 
The observed Solar System ratio of the europium isotopes in our
sample of three low-metallicity, old halo stars suggests that
successful {\it r}-process models will need to be those that match the 
Solar System {\it r}-process abundances, at least for the heaviest 
neutron-capture elements.
Whether there is a narrow range of astrophysical and nuclear conditions
(Cameron 2001\nocite{Cam01}),
a narrow mass range for the sites of the r-process (Mathews et~al. 
1992\nocite{MBC92}; Wheeler et~al. 1998\nocite{WCH98}), or some other
constraining factors, there appears to be a uniformity in the production
of these elements and, at least for europium, the isotopes.
Several groups have high resolution and S/N data for other very metal-poor
stars with enhanced rare-earth elemental abundances.
Further Eu isotopic abundance analyses should be undertaken
to determine if there are exceptions to the apparent equality
of Eu isotopic fractions in low metallicity halo stars.

\acknowledgments

We thank David Lambert and referee Glenn Wahlgren for helpful suggestions
that have improved our paper.
This research was funded in part by NSF grants 
AST-99987162 to CS, AST-9986974 to JJC, AST-9819400 to JEL,
AST-0098549 to TCB, and PHY-9800980 to GMF.

\clearpage

\clearpage

\clearpage

\end{document}